\documentclass[aps,amsmath,showpacs,showkeys,nofootinbib,
superscriptaddress]{revtex4}
\usepackage[dvips]{graphicx}

\begin{document}

\title{\bf Realistic calculations of nuclear disappearance lifetimes 
induced by $n\bar n$ oscillations} 

\author{E.~Friedman}
\email{elifried@vms.huji.ac.il}
\affiliation{Racah Institute of Physics, The Hebrew University,
Jerusalem 91904, Israel}

\author{A.~Gal}
\email{avragal@vms.huji.ac.il}
\affiliation{Racah Institute of Physics, The Hebrew University, 
Jerusalem 91904, Israel} 

\date{\today} 

\begin{abstract} 
Realistic calculations of nuclear disappearance lifetimes induced by 
$n\bar n$ oscillations are reported for oxygen and iron, using $\bar n$ 
nuclear potentials derived from a recent comprehensive analysis of 
$\bar p$ atomic X-ray and radiochemical data. A lower limit 
$\tau_{n\bar n} > 3.3 \times 10^8$~s on the $n\bar n$ oscillation 
time is derived from the Super-Kamiokande I new lower limit 
$T_d({\rm O})> 1.77 \times 10^{32}$~yr on the neutron lifetime in oxygen. 
Antineutron scattering lengths in carbon and nickel, needed in trap 
experiments using ultracold neutrons, are calculated from updated 
$\bar N$ optical potentials at threshold, with results shown to be 
largely model independent. 
\end{abstract}

\pacs{11.30.Fs, 13.75.Cs, 36.10.Gv}
 
\keywords{neutron-antineutron oscillations, neutron lifetimes, nuclear 
disappearance lifetimes, antiprotonic atoms, antineutron scattering lengths}  

\maketitle 


\section{Introduction and overview} 
\label{sec:intro} 

The stability of nuclei, as determined by looking for proton decay 
\cite{Tak86,Ber89,Chu02}, sets a lower limit also on the lifetime of 
other processes such as neutron-antineutron ($n \bar n$) oscillations 
in free space. This $\Delta B = 2$ baryon-violating $n \bar n$ 
oscillation process was pointed out long ago in the pioneering papers 
by Kuzmin \cite{Kuz70}, by Glashow \cite{Gla79} and by Mohaptra and 
Marshak \cite{MMa80}. Several quantitative calculations relating the 
nuclear disappearance lifetime $T_d$ to the $n \bar n$ oscillation 
time $\tau_{n \bar n}$ have been reported \cite{ABM82,DGR83,HKo98}. 
In these calculations a point-like $n \bar n$ coupling 
$\delta m = \hbar \tau_{n \bar n}^{-1}$ is assumed. In free space, 
$\delta m$ splits the $n - \bar n$ degenerate mass $m$ into mass eigenstates 
$m \pm \delta m$. The $n \bar n$ oscillations between these two mass 
eigenstates are suppressed in nuclear matter, giving place to decay of 
neutrons in a stable nucleus. Instead of two mass eigenstates one encounters 
two distinct widths, one which is the nuclear $\bar n$ annihilation width 
of order $\Gamma_{\bar n} \approx 320$~MeV for central nuclear densities 
\cite{FGM05}, the other one associated with the lifetime of a bound neutron: 
\begin{equation} 
\label{eq:supp} 
\Gamma_d \approx (\frac{4~\delta m}{\Gamma_{\bar n}})~\delta m~, 
\end{equation} 
where $\Gamma_d = \hbar T_d^{-1}$ is the nuclear disappearance width 
per neutron. These statements follow, more rigorously, from a discussion 
of the temporal evolution of nuclear disappearance owing to $n \bar n$ 
oscillations \cite{Gal00}. Eq.~(\ref{eq:supp}) demonstrates that 
$n \bar n$ oscillations are suppressed in nuclei by 31 orders of magnitude, 
which is the ratio between the time scales $10^{-23}$~s for the 
strong-interaction $\bar n$ annihilation time 
$\tau_{\bar n}=\hbar \Gamma_{\bar n}^{-1}$ and $10^8$~s presumed below for 
the free-space oscillation time $\tau_{n \bar n} = \hbar (\delta m)^{-1}$. 
In addition to $\bar n$ annihilation, 
neutrons and antineutrons also experience different nuclear potentials 
$U_n$ and $U_{\bar n}$, leading within a closure approximation \cite{CKK81} 
to a refinement of Eq.~(\ref{eq:supp}): 
\begin{equation} 
\label{eq:fullsupp} 
\Gamma_d^{\rm closure} \approx \Gamma_{\bar n}~\frac{(\delta m)^2}
{<W_{\bar n}>^2 + <(U_{\bar n}-U_n)>^2}~, 
\end{equation} 
where $<W_{\bar n}>=\Gamma_{\bar n}/2$ is some appropriately chosen average 
of (minus) the imaginary part of the $\bar n$ nuclear potential.  

To leading order, from Eq.~(\ref{eq:supp}), the relationship between 
$\tau_{n \bar n}$ and $T_d$ is given roughly by 
\begin{equation}
\tau_{n \bar n} \approx 2~(\hbar~T_{d}/\Gamma_{\bar n})^{1/2}~. 
\label{eq:relation} 
\end{equation} 
The values reported for $T_d$ in the literature, by convention, are normalized 
per neutron and take into account secondary and absorption processes. Thus, 
$T_d$ essentially stands for the lifetime of a neutron in a stable nucleus. 
Using the best value $T_d({\rm Fe}) > 7.2 \times 10^{31}$~yr published by the 
Soudan 2 Collaboration \cite{Chu02}, Eq.~(\ref{eq:relation}) gives 
a lower limit estimate of $\tau_{n \bar n} > 1.4 \times 10^8$~s. 
This estimate is to be compared with the lower limit 
$\tau_{n \bar n} > 1.3 \times 10^8$~s stated by the Soudan 2 
Collaboration using the calculations made long ago by Dover-Gal-Richard 
\cite{DGR83}. However, this apparent agreement might be fortuitous. 
We comment that the lower limit on $\tau_{n \bar n}$ derived from nuclear 
disappearance considerations is higher than that determined using 
nuclear-reactor neutrons directly in searches for $n \bar n$ oscillations. 
The lower limit given by the Grenoble reactor experiment \cite{Bal90} is 
$\tau_{n \bar n} > 0.86 \times 10^8$~s. 

Recently, the Super-Kamiokande (SK) collaboration released an improved value 
for $T_d$ in oxygen, $T_d(\rm O) > 1.77 \times 10^{32}$~yr \cite{SKI07}. 
In the present work we report an accurate calculation of the lower limit on 
$\tau_{n \bar n}$ implied by this value of $T_d(\rm O)$, using the latest 
detailed analysis by our group which derived antinucleon nuclear potentials 
from antiprotonic atom data and radiochemical data \cite{FGM05}. 
These potentials are essentially isoscalar and apply to antineutrons 
as well as to antiprotons. 
For completeness, in this study we also calculate the reduced lifetime 
(see below) in iron which will provide the necessary link between a future 
measurement of $T_d(\rm Fe)$ and a derivation of a lower limit on 
$\tau_{n \bar n}$. 

Another experimental method of looking for $n \bar n$ oscillations is to 
use ultracold neutrons in a trap, first suggested by Chetyrkin {\it et al.} 
\cite{CKK81} and discussed since then by several other groups, {\it e.g.} 
in Refs.~\cite{YGo92,KKL04}. A necessary input for interpreting such 
experiments are the nuclear $\bar n$ scattering lengths for ultracold 
antineutrons in matter. We report here on calculations of $\bar n$ scattering 
lengths for several chosen materials, again based on the analysis of 
antiprotonic atoms. The calculated values of these scattering lengths turn 
out to be largely model independent owing to the strong absorption of low 
energy antinucleons.

\section{Methodology} 
\label{sec:meth} 

\subsection{Wave equations, widths and lifetime} 

A point-like coupling $\delta m = \hbar \tau_{n \bar n}^{-1}$, representing 
$n \bar n$ oscillations in free space, connects each of the stationary 
bound single-particle (sp) neutron states to the corresponding $\bar n$ 
stationary sp state. The sp energies assume complex values $E_{\nu{\ell}j}
=-B_{\nu{\ell}j}-{\rm i}{\Gamma_{\nu{\ell}j}}/2$, where the imaginary 
part of the energy in the sp state labeled ${\nu{\ell}j}$ gives the 
disappearance width $\Gamma_{\nu{\ell}j}$ for this state. 
The coupled $n$-$\bar n$ radial wave equations for the neutron sp 
wavefunction $u_{\nu{\ell}j}(r)$ and the antineutron sp 
wavefunction $w_{\nu{\ell}j}(r)$ are given by 

\begin{equation} 
-\frac{{\hbar}^2}{2\mu}~u_{\nu{\ell}j}^{''}(r)+\frac{{\hbar}^2
{\ell}({\ell}+1)}{2\mu r^2}~u_{\nu{\ell}j}(r)-U_n(r)~u_{\nu{\ell}j}(r)
-E_{\nu{\ell}j}~u_{\nu{\ell}j}(r)+\delta m~w_{\nu{\ell}j}(r)=0~, 
\label{eq:n} 
\end{equation} 
\begin{equation} 
-\frac{{\hbar}^2}{2\mu}~w_{\nu{\ell}j}^{''}(r)+\frac{{\hbar}^2{\ell}({\ell}+1)}
{2\mu r^2}~w_{\nu{\ell}j}(r)-[U_{\bar n}(r)+{\rm i}~W_{\bar n}(r)]~w_{\nu{\ell}
j}(r)-E_{\nu{\ell}j}~w_{\nu{\ell}j}(r)+\delta m~u_{\nu{\ell}j}(r)=0~, 
\label{eq:nbar} 
\end{equation} 
where $-U_n(r)$ and $-(U_{\bar n}(r)+{\rm i}~W_{\bar n}(r))$ are the nuclear 
potentials exerted by the nuclear core on the neutrons and antineutrons, 
respectively, and $\mu$ is the reduced mass. 
The radial wavefunctions $u_{\nu{\ell}j}(r)$ and $w_{\nu{\ell}j}(r)$ are 
regular at the origin and decay with $r$ outside of the nucleus. A useful 
expression for the width $\Gamma_{\nu{\ell}j}$ is obtained by multiplying 
Eq.~(\ref{eq:n}) by $u_{\nu{\ell}j}^{*}(r)$, and the complex conjugate of 
Eq.~(\ref{eq:n}) by $u_{\nu{\ell}j}(r)$, subtracting the resulting expressions 
from each other and integrating from zero to infinity. The result is 
\begin{equation} 
\Gamma_{\nu{\ell}j} = - \frac{2~\delta m~
\int{{\rm Im}~(w_{\nu{\ell}j}(r)~u_{\nu{\ell}j}^{*}(r))~dr}}
{\int{\mid{u_{\nu{\ell}j}(r)}\mid}^2~dr}~. 
\label{eq:width} 
\end{equation} 
The initial-time boundary condition of no antineutrons implies that 
$\mid{w/u}\mid = O({\delta m}/B)$, where the binding energy $B$ represents any 
of the strong-interaction entities, such as $\Gamma_{\bar n}$ \cite{Gal00}. 
It follows then from Eq.~(\ref{eq:width}) that the width $\Gamma$ is of order 
$(\delta m)^2/\Gamma_{\bar n}$, in agreement with Eq.~(\ref{eq:supp}). 
The terms with $\Gamma$ and $\delta m$ in Eq.~(\ref{eq:n}) are negligible, 
of order $({\delta m}/B)^2$ with respect to the rest of the terms which 
coincide with those constituting a stable bound-neutron radial wave equation 
in which $B_{\nu{\ell}j}=B_{\nu{\ell}j}^{(n)}$ stands for the neutron sp 
binding energy. Thus, the solutions $u_{\nu{\ell}j}$ are 
essentially real functions. Eq.~(\ref{eq:width}) expresses a relationship 
between these two minute terms which are neglected below. 
Turning to Eq.~(\ref{eq:nbar}), all terms there are of the same order, 
except for the $\Gamma$ term which is of order $({\delta m}/B)^2$ with 
respect to the other terms and, hence, may be dropped off. 

Dropping off the terms of order $({\delta m}/B)^2$ in Eqs.~(\ref{eq:n}) and 
(\ref{eq:nbar}), we obtain the radial equations satisfied (i) by 
a bound-neutron wavefunction, 
\begin{equation} 
-\frac{{\hbar}^2}{2\mu}u_{\nu{\ell}j}^{''}(r)+\frac{{\hbar}^2{\ell}({\ell}+1)}
{2\mu r^2}u_{\nu{\ell}j}(r)-U_n(r)u_{\nu{\ell}j}(r)
+B_{\nu{\ell}j}^{(n)}~u_{\nu{\ell}j}(r)=0~, 
\label{eq:nfinal} 
\end{equation} 
and (ii) by a quasibound antineutron {\it reduced} wavefunction 
$v_{\nu{\ell}j}(r)=w_{\nu{\ell}j}(r)/\delta m$, 
\begin{equation} 
-\frac{{\hbar}^2}{2\mu}v_{\nu{\ell}j}^{''}(r)+\frac{{\hbar}^2{\ell}({\ell}+1)}
{2\mu r^2}v_{\nu{\ell}j}(r)-[U_{\bar n}(r)+{\rm i}~W_{\bar n}(r)]
v_{\nu{\ell}j}(r)+
B_{\nu{\ell}j}^{(n)}~v_{\nu{\ell}j}(r)+u_{\nu{\ell}j}(r)=0~. 
\label{eq:nbarfinal} 
\end{equation} 
Operating on Eq.~(\ref{eq:nbarfinal}) similarly to the way in which 
Eq.~(\ref{eq:width}) was derived from Eq.~(\ref{eq:n}), recalling that 
$B_{\nu{\ell}j}^{(n)}$ and $u_{\nu{\ell}j}(r)$ are real, and multiplying 
the result by $(\delta m)^2$ in order to make connection with 
Eq.~(\ref{eq:width}), we obtain 
\begin{equation} 
-2~(\delta m)^2 \int{{\rm Im}~(v_{\nu{\ell}j}(r)~u_{\nu{\ell}j}^{*}(r))~dr}=
2~(\delta m)^2 \int{W_{\bar n}(r)~{\mid{v_{\nu{\ell}j}(r)}\mid}^2~dr}~, 
\label{eq:test} 
\end{equation} 
so that the disappearance width from the ${\nu{\ell}j}$ sp state, 
Eq.~(\ref{eq:width}), is given by    
\begin{equation} 
\Gamma_{\nu{\ell}j}=\frac{2~(\delta m)^2 \int{W_{\bar n}(r)~{\mid{v_{\nu{\ell}
j}(r)}\mid}^2~dr}}{\int{u_{\nu{\ell}j}^2(r)~dr}}=-\frac{2~(\delta m)^2 \int{u_
{\nu{\ell}j}(r)~{\rm Im}~v_{\nu{\ell}j}(r)~dr}}{\int{u_{\nu{\ell}j}^2(r)~dr}} 
\label{eq:partialGamma} 
\end{equation} 
in terms of the solutions $u_{\nu{\ell}j}$ and $v_{\nu{\ell}j}$ of 
Eqs.~(\ref{eq:nfinal}) and (\ref{eq:nbarfinal}), respectively. 
The averaged disappearance width {\it per neutron} is then given by 
\begin{equation} 
\Gamma_d = \frac{1}{N}\sum{n_{\nu{\ell}j} \Gamma_{\nu{\ell}j}}~, 
\label{eq:Gamma} 
\end{equation} 
where $n_{\nu{\ell}j}$ is the appropriate number of neutrons in the sp 
state $\nu{\ell}j$, $N=\sum{n_{\nu{\ell}j}}$ is the number of neutrons 
in the decaying nucleus and summation is over the occupied neutron 
sp states. Since $\Gamma_d$ scales as $(\delta m)^2$, hence inversely 
proportional to $\tau_{n \bar n}^2$, it is customary to define 
a {\it reduced lifetime} $T_R$ given by 
\begin{equation} 
T_R = \frac{\hbar}{\Gamma_d~\tau_{n \bar n}^2}~, 
\label{eq:Treduced} 
\end{equation} 
which has dimension of inverse time (s$^{-1}$). 
The nuclear disappearance lifetime $T_d$ is then given by 
\begin{equation} 
T_d = \frac{\hbar}{\Gamma_d}=T_R~\tau_{n \bar n}^2~.  
\label{eq:Td} 
\end{equation} 

We solve numerically both Eqs.~(\ref{eq:nfinal}) and (\ref{eq:nbarfinal}) 
for neutron and antineutron sp states, respectively. Eq.~(\ref{eq:nfinal}) 
is identical with that used in nuclear bound state problems for occupied sp 
neutron states. Eq.~(\ref{eq:nbarfinal}) is solved for each one of the 
antineutron sp states with reduced wavefunctions $v_{\nu{\ell}j}$ that are 
generated by the corresponding occupied sp neutron wavefunctions 
$u_{\nu{\ell}j}$ acting as an inhomogeneous source. The solutions 
$u_{\nu{\ell}j}$ and $v_{\nu{\ell}j}$ serve as input in the integrals 
Eq.~(\ref{eq:partialGamma}) for the widths $\Gamma_{\nu{\ell}j}$. 
For completeness we note the precise expression for $\Gamma_{\nu{\ell}j}$, 
without neglecting contributions of order $({\delta m}/B)^2$: 
\begin{equation} 
\Gamma_{\nu{\ell}j}=\frac{2\int{W_{\bar n}(r)~{\mid{w_{\nu{\ell}j}(r)}
\mid}^2~dr}}{\int{({\mid{u_{\nu{\ell}j}(r)}\mid}^2+
{\mid{w_{\nu{\ell}j}(r)}\mid}^2)~dr}}~, 
\label{eq:complete} 
\end{equation} 
where $u_{\nu{\ell}j}$ and $w_{\nu{\ell}j}$ solve Eqs.~(\ref{eq:n}) and 
(\ref{eq:nbar}).

\subsection{Nuclear structure issues} 

The standard shell-model (SM) description of neutron sp states in nuclei 
introduces spurious excitations of the center of mass degree of freedom. It 
is important, particularly in as light a nucleus as $^{16}$O, to eliminate 
this spuriosity and thus avoid its unphysical effects on the equations solved 
for the neutron sp states and on the neutron disappearance widths subsequently 
derived. A general construction of translationally invariant (TI) nuclear 
wavefunctions and densities, in the harmonic-oscillator basis, was given by 
Navr\'{a}til \cite{Nav04}. Here we follow the earlier discussion by Millener 
{\it et al.} \cite{MOW83} which is specifically suited to the $p$ shell. 
Solving radial equations in the relative coordinate between a neutron and its 
nuclear core in a $p$-shell nucleus, the number of neutrons in the $s$ shell 
and $p$ shell has to be modified from the SM values $n^{\rm SM}_{1s}=2$ and 
$n^{\rm SM}_{1p}=(N-2)$ to 
\begin{equation} 
\label{nWS} 
n^{\rm TI}_{1s}=2-\frac{N-2}{A-1}~,~~~~ n^{\rm TI}_{1p}={\frac{A}{A-1}}(N-2)~, 
\end{equation} 
where for $N=8$ appropriate to $^{16}$O we have $n^{\rm TI}_{1s}=1.6$ and 
$n^{\rm TI}_{1p}=6.4$. To reproduce a given value of the mean-square (ms) 
radius of the point-neutron distribution $<r^2>_n$, the ms radii of the 
neutron-core $1s$ and $1p$ wavefunctions have to satisfy 
\begin{equation} 
\label{eq:msTI} 
<r^2>_n={\frac{1}{N}}(\frac{A-1}{A})^2~(n^{\rm TI}_{1s}<r^2>_{1s}
+~n^{\rm TI}_{1p}<r^2>_{1p})~, 
\end{equation} 
where it was assumed that the $s$-hole and $p$-hole strengths are not 
fragmented. In practice we used a spin-orbit potential to split the 
$p$-hole strength according to the observed $p_{1/2}-p_{3/2}$ energy 
difference. Equation~(\ref{eq:msTI}) is to be contrasted with the 
SM version in which center of mass spuriosities are disregarded:  
\begin{equation} 
\label{eq:msSM} 
<r^2>_n={\frac{1}{N}}~(n^{\rm SM}_{1s}<r^2>_{1s}
+~n^{\rm SM}_{1p}<r^2>_{1p})~. 
\end{equation}

\subsection{Numerical solution} 

The real wavefunction $u_{\nu{\ell}j}$ for the bound neutron is obtained 
by solving numerically Eq. (\ref{eq:nfinal}) using a standard method. Here 
we describe briefly the method used to solve the inhomogenous equation 
(\ref{eq:nbarfinal}) for the quasibound antineutron reduced wavefunction 
in the potential taken from fits to antiprotonic atom data \cite{FGM05}. 

The inhomogeneous radial equation (\ref{eq:nbarfinal}) is integrated 
numerically from $r=0$ outward using the Numerov method, requiring the 
usual regular boundary conditon $v_{\nu{\ell}j} \sim r^{\ell +1}$ at the 
origin. In parallel, the corresponding homogeneous radial equation 
\begin{equation} 
-\frac{{\hbar}^2}{2\mu}{v_{\nu{\ell}j}^{(0)}}^{''}(r)+\frac{{\hbar}^2{\ell}
({\ell}+1)}{2\mu r^2}v_{\nu{\ell}j}^{(0)}(r)-[U_{\bar n}(r)+{\rm i}~W_{\bar n}
(r)]v_{\nu{\ell}j}^{(0)}(r)+B_{\nu{\ell}j}^{(n)}~v_{\nu{\ell}j}^{(0)}(r)=0~, 
\label{eq:w} 
\end{equation} 
obtained from Eq.~(\ref{eq:nbarfinal}) by omitting the last term, 
is also integrated using the regular boundary condition at $r=0$.
Integration is carried out to a matching radius $R$ where the 
nuclear $\bar n$ potentials may safely be neglected. Both integrations 
lead to exponentially increasing functions toward $R$, as expected. 
The most general, regular at $r=0$ solution of Eq.~(\ref{eq:nbarfinal}) 
is given by the linear combination 
\begin{equation} 
v_{\nu{\ell}j}^<~ =~ a_{\nu{\ell}j}~v_{\nu{\ell}j}^{(0)} + v_{\nu{\ell}j}~, 
\label{eq:combin} 
\end{equation} 
where the (complex) constant $a_{\nu{\ell}j}$ is chosen such that 
$v_{\nu{\ell}j}^<$ is regular also at infinity. 

We note that outside the matching radius $R$ the homogeneous equation 
(\ref{eq:w}) is satisfied by the neutron bound-state wavefunction 
$u_{\nu{\ell}j}$. The most general, regular at $r \rightarrow \infty$ 
solution of Eq.~(\ref{eq:nbarfinal}) is then given by the linear combination 
\begin{equation} 
v_{\nu{\ell}j}^>~=~ b_{\nu{\ell}j}~u_{\nu{\ell}j} + {\tilde v}_{\nu{\ell}j}~, 
\label{eq:combout} 
\end{equation} 
where ${\tilde v}_{\nu{\ell}j}$ is a special, regular at 
$r \rightarrow \infty$ solution of the inhomogeneous equation 
(\ref{eq:nbarfinal}), and $b_{\nu{\ell}j}$ is an arbitrary (complex) constant. 

The constants $a_{\nu{\ell}j}$ and $b_{\nu{\ell}j}$ above are determined 
by requiring that the `inside' and `outside' solutions, $v_{\nu{\ell}j}^<$ 
and $v_{\nu{\ell}j}^>$ respectively, as well as their first derivatives, 
agree with each other at the matching point $r=R$. We note that for the 
purpose of evaluating the disappearance widths $\Gamma_{\nu{\ell}j}$, 
only the knowledge of the constants $a_{\nu{\ell}j}$ is required. 
A straightforward algebra gives  
\begin{equation} 
a_{\nu{\ell}j}~=~\frac{[u_{\nu{\ell}j}'(R){\tilde v}_{\nu{\ell}j}(R) - 
u_{\nu{\ell}j}(R){\tilde v}_{\nu{\ell}j}'(R)]~-~
[u_{\nu{\ell}j}'(R)v_{\nu{\ell}j}(R)-u_{\nu{\ell}j}(R)v_{\nu{\ell}j}'(R)]}
{u_{\nu{\ell}j}'(R)v_{\nu{\ell}j}^{(0)}(R)-
u_{\nu{\ell}j}(R){v_{\nu{\ell}j}^{(0)}}'(R)}~.
\label{eq:boundary} 
\end{equation} 
Since both $u_{\nu{\ell}j}$ and ${\tilde v}_{\nu{\ell}j}$ fall off 
exponentially, whereas $v_{\nu{\ell}j}$ increases exponentially at $R$, the 
terms in the first square bracket in the numerator of Eq.~(\ref{eq:boundary}) 
are of order $\exp(-2\kappa_{\nu{\ell}j}R)$ with respect to the terms in 
the second square bracket, and may be safely neglected. 
Here $\kappa_{\nu{\ell}j}=(2\mu~B_{\nu{\ell}j}^{(n)})^{1/2}/{\hbar}$, and 
$\exp(-2\kappa_{\nu{\ell}j}R)<10^{-4}$ for a wide range of realistic 
nuclear applications. Hence, to such high accuracy 
\begin{equation} 
a_{\nu{\ell}j}~\approx~ - \frac{u_{\nu{\ell}j}'(R)v_{\nu{\ell}j}(R) - 
u_{\nu{\ell}j}(R)v_{\nu{\ell}j}'(R)}
{u_{\nu{\ell}j}'(R)v_{\nu{\ell}j}^{(0)}(R)-
u_{\nu{\ell}j}(R){v_{\nu{\ell}j}^{(0)}}'(R)}~, 
\label{eq:aa} 
\end{equation}
and no specific knowledge of the special solutions ${\tilde v}_{\nu{\ell}j}$ 
is required. [Special solutions ${\tilde v}_{\nu{\ell}j}$ in terms of 
neutron wavefunctions $u_{\nu{\ell}j}$ are given in the Appendix.] 
A further simplification of Eq.~(\ref{eq:aa}) occurs by noting that 
$v_{\nu{\ell}j}$ in the numerator and $v_{\nu{\ell}j}^{(0)}$ in the 
denominator, for any given $\ell$ value, share the same asymptotic behavior 
at $R$, which leads to the final result 
\begin{equation} 
a_{\nu{\ell}j}~\approx~-\frac{v_{\nu{\ell}j}(R)}{v_{\nu{\ell}j}^{(0)}(R)}~.
\label{eq:aafinal} 
\end{equation} 
These expressions are useful only when their dependence on the matching 
radius $R$ is negligible. In practice we used Eq.~(\ref{eq:aa}) with 
radial integration steps of 0.04~fm and $R$ between 10 and 13 fm. The 
coefficients $a_{\nu{\ell}j}$ which determine the required decaying 
$\bar n$ radial wavefunctions in the region $r<R$ were found to vary by 
less than $10^{-5}$ of their values, with the resulting widths stable to 
better than $10^{-4}$. Both forms of Eq.~(\ref{eq:partialGamma}) gave 
identical results to this order. The same accuracy is also obtained using 
Eq.~(\ref{eq:aafinal}). For comparison with previous calculations, 
we calculated the widths for $^{16}$O, using the input paprameters given 
by Dover {\it et al.} \cite{DGR83}. The results agreed with those listed 
in Table I of that reference to within $1\%$.  

\begin{figure} 
\includegraphics[scale=0.56,angle=0]{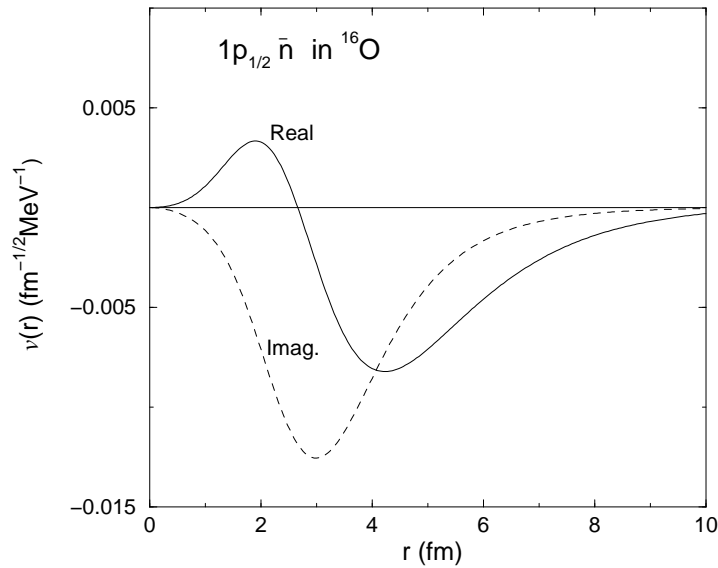} 
\caption{Antineutron $1p_{1/2}$ reduced radial wavefunction 
$v$ in $^{16}$O.} 
\label{fig:fglblfig1} 
\end{figure} 

\begin{figure} 
\includegraphics[scale=0.56,angle=0]{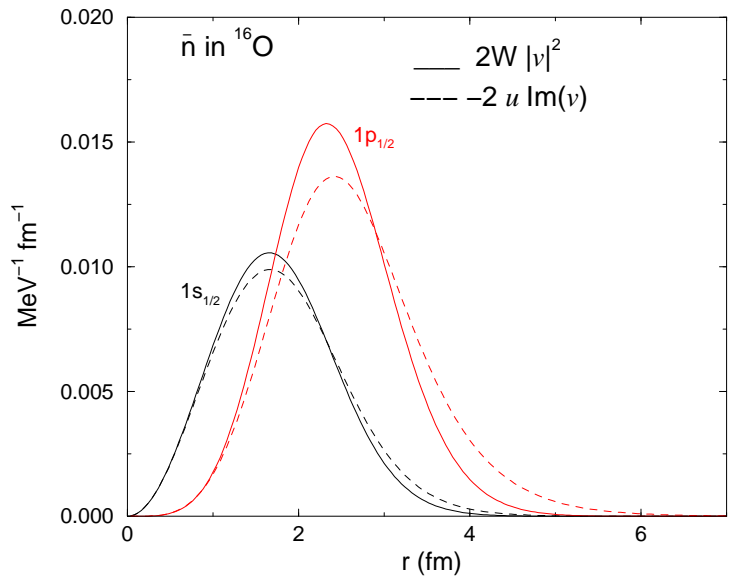} 
\caption{Integrands in the numerators for the two versions 
of the neutron disappearance width Eq.~(\ref{eq:partialGamma}).} 
\label{fig:fglblfig2} 
\end{figure} 

Figure~\ref{fig:fglblfig1} shows, as an example, the reduced antineutron 
$1p_{1/2}$ wavefunction $v$ in $^{16}$O. Note that the longer tail 
of the real part is of no consequence for the disappearance width 
because in the ${\mid{v_{\nu{\ell}j}(r)}\mid}^2$ version of the 
integral, Eq.~(\ref{eq:partialGamma}), the imaginary potential $W$ 
is of a shorter range. The other version of the width contains only 
the imaginary part of the antineutron wavefunction, which is of 
a shorter range as is seen in the figure. Figure~\ref{fig:fglblfig2} 
shows the integrands in the numerator for the two versions of the width, 
for $1s_{1/2}$ and $1p_{1/2}$ neutrons in $^{16}$O. It is remarkable 
that the relative differences between the integrals are less than $10^{-5}$, 
although the two integrands are not identical both 
near the maxima and in the tail region. 

\begin{table} 
\caption{Calculated reduced neutron disappearance widths 
$\Gamma_{\nu{\ell}j}/(\delta m)^2$ (in MeV$^{-1}$).}  
\label{tab:widths} 
\begin{ruledtabular} 
\begin{tabular}{lccccc} 
state $\nu{\ell}j$ & $n^{\rm SM}_{\nu{\ell}j}$ & 
$n^{\rm TI}_{\nu{\ell}j}(^{16}{\rm O})$ & $^{16}$O(SM) & $^{16}$O(TI)& 
$^{56}$Fe(SM) \\ 
\hline  
 1$s_{1/2}$ & 2 & 1.6000 & 0.0207 & 0.0220 & 0.0144 \\ 
 1$p_{3/2}$ & 4 & 4.2667 & 0.0296 & 0.0296 & 0.0177 \\ 
 1$p_{1/2}$ & 2 & 2.1333 & 0.0321 & 0.0343 & 0.0177 \\ 
 1$d_{5/2}$ & 6 & $-$      & $-$      & $-$      & 0.0217 \\ 
 2$s_{1/2}$ & 2 & $-$      & $-$      & $-$      & 0.0238 \\  
 1$d_{3/2}$ & 4 & $-$      & $-$      & $-$      & 0.0219 \\  
 1$f_{7/2}$ & 8 & $-$      & $-$      & $-$      & 0.0268 \\ 
 2$p_{3/2}$ & 2 & $-$      & $-$      & $-$      & 0.0343 \\  \hline 
 $\Gamma_d/(\delta m)^2$-average [Eq.~(\ref{eq:Gamma})] &&& 0.0280 & 0.0294 
 & 0.0228 \\
 $\Gamma_d/(\delta m)^2$-closure [Eq.~(\ref{eq:fullsupp})] &&& 0.0271 & 0.0265 
 & 0.0220 \\ 
 $T_R$ [Eq.~(\ref{eq:Treduced}] (s$^{-1}$) &&& 0.543 $\times 10^{23}$ & 
 0.517 $\times 10^{23}$ & 0.666 $\times 10^{23}$ \\
\end{tabular} 
\end{ruledtabular} 
\end{table}

\section{Results} 
\label{sec:res}

\subsection{Neutron disappearance widths} 
\label{sec:gammares} 

\begin{figure} 
\includegraphics[scale=0.56,angle=0]{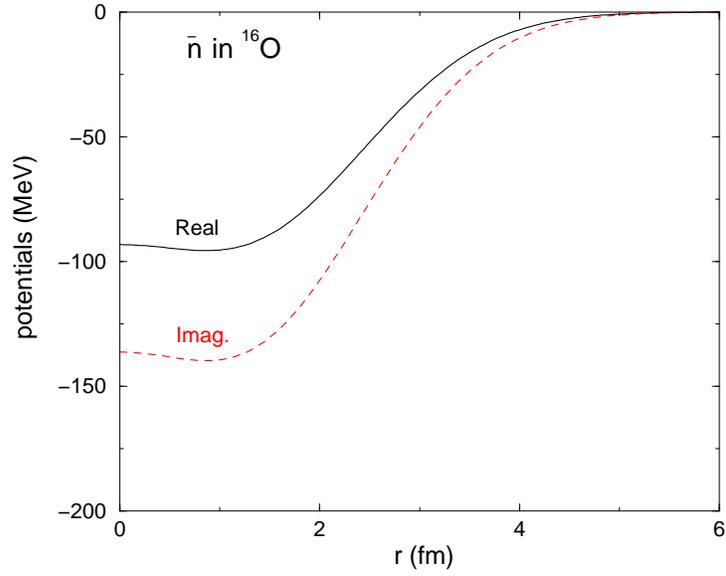} 
\caption{The antineutron optical potential in $^{16}$O.} 
\label{fig:fglblfig3} 
\end{figure}

\begin{figure} 
\includegraphics[scale=0.56,angle=0]{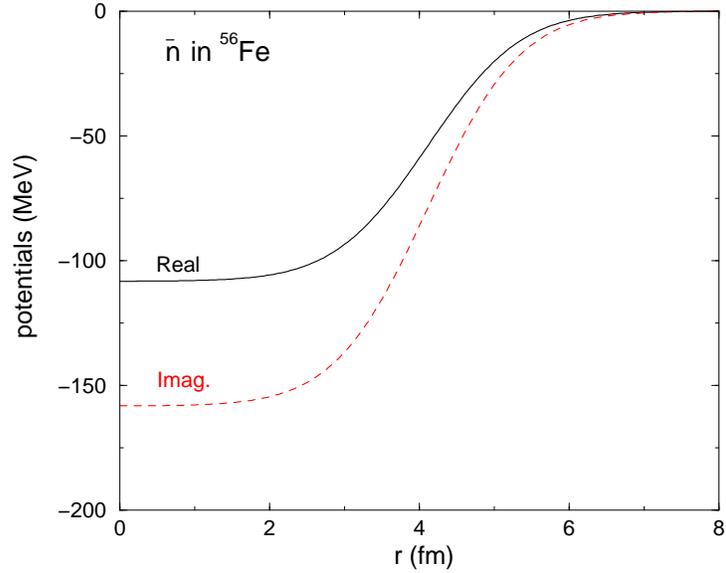} 
\caption{The antineutron optical potential in $^{56}$Fe.} 
\label{fig:fglblfig4} 
\end{figure}

Neutron disappearance widths for the various subshells $\nu{\ell}j$ 
were calculated for $^{16}$O and $^{56}$Fe with the two forms given 
by Eq.~(\ref{eq:partialGamma}). The bound neutron wavefunctions 
$u_{\nu{\ell}j}$ were calculated in a Woods-Saxon potential 
~$-U_0^{(n)}/[1 + \exp((r-R_{1/2})/a)]$ whose depth $U_0^{(n)}$ 
was adjusted in each nucleus to fit the experimental separation energy 
of the least-bound neutron. A spin-orbit term was added to reproduce the 
observed $p_{1/2}-p_{3/2}$ splitting. The half-density radius parameters 
$R_{1/2}=r_0(A-1)^{1/3}$ were adjusted such that the root-mean-square (rms) 
radius of the whole neutron distributions was 2.57 fm for $^{16}$O and 3.71 
fm for $^{56}$Fe. For $^{16}$O it corresponds to the known rms radius for 
the point-proton distribution and for $^{56}$Fe it is 0.09 fm larger than the 
known value for the point-proton distribution \cite{FBH95}. The diffusivity 
parameter $a$ was fixed at $a=0.60$~fm for $^{16}$O and $a=0.55$~fm for 
$^{56}$Fe. Using Eq.~(\ref{eq:msSM}) for $^{16}$O and its straightforward 
extension for $^{56}$Fe, values of $r_0=1.325$~fm and $r_0=1.304$~fm were 
found for the SM calculations in $^{16}$O and $^{56}$Fe, respectively, whereas 
using Eq.~(\ref{eq:msTI}) a value of $r_0=1.442$~fm was found for the TI 
calculation in $^{16}$O. The depths of the SM potential were 53.8 MeV for 
$^{16}$O and 51.1 MeV for $^{56}$Fe, and the depth of the TI potential in 
$^{16}$O was 48.8 MeV. For the antineutrons we used 
the most recent phenomenological isoscalar potential obtained from 
large-scale (`global') fits to 90 data points of strong-interaction shifts 
and widths in antiprotonic atoms across the periodic table \cite{FGM05}. 
The effective amplitude for that potential is $b_0=1.3+{\rm i}~1.9$ fm, 
used with a finite-range Gaussian folded with a range parameter 0.9 fm 
into the nuclear matter density; see Ref.~\cite{FGM05} for more details. 
These potentials are shown in Fig.~\ref{fig:fglblfig3} and 
Fig.~\ref{fig:fglblfig4} for ${\bar n} - ^{15}$O and ${\bar n} - ^{55}$Fe, 
respectively. 

The neutron disappearance widths as given by Eq.~(\ref{eq:partialGamma}) 
scale with $(\delta m)^2$. Therefore in Table \ref{tab:widths} we list 
the calculated reduced neutron disappearance widths 
$\Gamma_{\nu{\ell}j}/(\delta m)^2$ for the various sub-shells in 
$^{16}$O and in $^{56}$Fe. These calculated widths increase by as much as 
$50\%$ in $^{16}$O and by over $100\%$ in $^{56}$Fe going outward from the 
inner $1s_{1/2}$ sp neutron state to the least-bound sp neutron state, in 
agreement with the trend found in earlier calculations \cite{DGR83}.  
The values of $\Gamma_d/(\delta m)^2$ 
calculated using the closure expression Eq.~(\ref{eq:fullsupp}) 
give excellent approximation to the exact weighted averages, provided 
values of the nuclear potentials $U_n,~U_{\bar n},~W_{\bar n}$ at the 
nuclear surface, here approximated by half of the corresponding values 
at the center of the nucleus, are adopted. We comment that for the strongly 
absorptive $\bar n$ potential used by us, the effect of the real potentials 
$U_n$ and $U_{\bar n}$ is secondary to that of the imaginary potential 
$W_{\bar n}$. A more elaborate averaging with the 
nuclear density should give similar results for strongly absorptive 
$\bar n$ potentials, as found by Dover {\it et al.} \cite{DGR83}. 
As for using the TI or the SM schemes for $^{16}$O, the difference between 
the averaged widths, or between the reduced lifetimes, amounts to merely 
$5\%$ which affects the limit placed on $\tau_{n\bar n}$ to only $2.5\%$. 
Our calculated values of the reduced lifetimes $T_R$ are smaller than 
those calculated in Ref.~\cite{DGR83}. This difference reflects partly the 
difference in the $\bar n$ potentials used and partly the more precise 
treatment of the nuclear geometry in our calculations. The half-density 
radius parameter of the WS potential used by Dover {\it et al.} in the SM 
calculation for $^{16}$O was taken as $R_{1/2}=2.545$~fm, whereas it assumes 
the value $R_{1/2}=3.268$~fm in the present SM calculation for $^{16}$O. 
The values of $R_{1/2}$ used for $^{56}$Fe are very close to each other in 
these works. Finally, for the recently reported SK result,  
$T_d(\rm O) > 1.77 \times 10^{32}$~yr \cite{SKI07} in oxygen, our calculated 
value for the reduced neutron disappearance width, using the TI version, 
implies a lower limit of $\tau_{n\bar n} > 3.3 \times 10^8$~s.

\subsection{Antineutron scattering lengths} 

For ultracold neutron experiments searching for $n \bar n$ oscillations, 
the knowledge of their scattering lengths in a given material is essential 
for constructing the relevant Fermi pseudopotentials. The neutron scattering 
lengths are known from thermal neutron reactions. Here we discuss the 
extraction of antineutron scattering lengths from the $\bar p$ optical 
potentials determined from a comprehensive analysis of $\bar p$ atomic 
data. We note that for the best-fit $\bar p$ potentials, the isovector 
component came out negligible \cite{FGM05}, so the $\bar n$ and $\bar p$ 
potentials are identical. Antinucleon scattering lengths were discussed 
extensively in Ref.~\cite{BFG01}. It was found there that a simple global 
parameterization was possible, in the form 
\begin{equation} 
{\rm Re}~a = (1.54 \pm 0.03)A^{0.311 \pm 0.005} \; {\rm fm}~ , \;\;\; 
{\rm Im}~a = -1.00 \pm 0.04 \; {\rm fm}~,  
\label{eq:a} 
\end{equation} 
for $A > 10$. The approximate $A^{1/3}$ dependence of the real part, 
and the constancy of imaginary part, are to be expected on the basis 
of a simple model based on a strongly absorptive square well potential, 
although the actual magnitude of Im $a$ is considerably larger than 
expected for a sharp-edge potential, resulting mainly from the 
diffuseness of the potential \cite{Bat83}. 

\begin{table}
\caption{$\bar n$ scattering lengths $a$ (in fm) from fits to $\bar p$ 
atomic data.} 
\label{tab:a} 
\begin{ruledtabular} 
\begin{tabular}{cccccc} 
Nucleus & FR~\cite{FGM05} & ZR~\cite{FGM05} & Eq.~(\ref{eq:a}) & 
S~\cite{WKS84} & D~\cite{WKS84} \\ 
\hline 
$^{12}$C & $3.26-{\rm i}~0.97$ & $3.21-{\rm i}~0.84$ & $(3.34 \pm 0.07)
-{\rm i}~(1.00 \pm 0.04)$ & $3.16-{\rm i}~0.87$ & $3.11-{\rm i}~1.10$ \\
$^{58}$Ni & $5.41-{\rm i}~1.12$ & $5.43-{\rm i}~1.14$ & $(5.44 \pm 0.11)
-{\rm i}~(1.00 \pm 0.04)$ & & \\ 
\end{tabular} 
\end{ruledtabular}
\end{table} 

In our latest work \cite{FGM05} the data base has been extended to include 
the numerous CERN PS209 Collaboration $\bar p$ atomic data. The values of 
$a$ due to the isoscalar $\bar N$ potentials fitted to this extended set 
of data are listed for $^{12}$C and $^{58}$Ni in Table~\ref{tab:a}, 
together with values obtained using Eq.~(\ref{eq:a}), and also from 
the earlier work by Wong {\it et al.} \cite{WKS84}. Here, FR and ZR stand 
for values of $a$ derived from best global-fit finite-range and zero-range 
potentials, respectively, with FR giving the lowest $\chi^2$ \cite{FGM05}. 
The notations S (shallow) and D (deep) stand for values of $a$ calculated 
in Ref.~\cite{YGo92} from potentials with a relatively weak absorptivity 
$W$ or a strong one, respectively, derived from limited fits in 
Ref.~\cite{WKS84}. The resulting scattering lengths $a$ exhibit 
a remarkable independence of the model used, provided it fits the $\bar p$ 
atomic data. This stability follows from the strong absorptivity of the 
$\bar p$ potential which suppresses the associated $1s$ atomic radial 
wavefunction in the nuclear interior where the main model dependence 
arises \cite{FGa99}.

\section{Summary} 
\label{sec:disc} 

We have reported results of precise calculations of the reduced lifetimes 
of representative nuclei, $^{16}$O and $^{56}$Fe, against neutron-antineutron 
oscillations, thus providing revised and updated lower limits on the 
free-space $n \bar n$ oscillation time $\tau_{n \bar n}$. The best lower 
limit is now provided by the very recent SK measurement \cite{SKI07} in 
$^{16}$O which yields according to our calculation a limit of 
$\tau_{n\bar n} > 3.3 \times 10^8$~s. We have used the latest (isoscalar) 
antinucleon potentials derived from the analysis of a large-scale set 
of $\bar p$ atomic data near threshold \cite{FGM05}. Having solved accurately 
the sp equations for neutrons and (coupled) antineutrons, it became possible 
to test the usefulness of rough approximations such as Eq.~(\ref{eq:fullsupp}) 
in terms of mean nuclear potentials for neutrons and antineutrons. We found 
that using surface values for these mean potentials, taken as half the 
corresponding values in the center of the nucleus, provided an excellent 
approximation to the exact calculation. This points out to a considerably 
mild model dependence of the calculated reduced widths that are sensitive 
foremost to $\bar n$ potentials at the nuclear surface where their 
determination from $\bar p$ atomic data involves only little extrapolation. 

An educated estimate of the {\it theoretical} uncertainty involved in the 
derivation of the lower bound deduced in the present work on the 
$n \bar n$ oscillation time $\tau_{n\bar n}$ can be made as follows. 
\begin{itemize} 
\item For a given nucleus like $^{16}$O and a given $\bar n$-nuclear 
potential, but considering alternative ways of treating the nuclear size, 
the calculated nuclear widths vary by $5\%$ (columns 4 and 5 of 
Table~\ref{tab:widths}), so the uncertainty in the derived $\tau_{n\bar n}$ 
is about $2.5\%$.
\item The uncertainty arising from using different nuclei (columns 4 and 6 
of Table~\ref{tab:widths}) comes mostly from the uncertainty in the strength 
of the absorptive (imaginary) $\bar n$-nuclear potential. That shows about 
$20\%$ uncertainty for the averaged disappearance width $\Gamma_d$, 
and hence $10\%$ for $\tau_{n\bar n}$. 
\end{itemize} 
Thus, the overall theoretical uncertainty involved in the present 
{\it one-nucleon} $n \bar n$ oscillation calculations is about $10\%-15\%$. 
It should be viewed as a model-dependence uncertainty that is considerably 
lower than the $50\%-100\%$ uncertainty range evident in many of the 
calculations from the 1980s and 1990s, {\it e.g.} Ref.~\cite{DGR83}, 
before the information from $\bar p$ atoms became as abundant and precise 
as it is available to date \cite{FGM05}. Other past calculations 
\cite{ABM82,HKo98} which avoided using $\bar n$ phenomenological optical 
potentials faced a tougher task of having to renormalize the $\bar n N$ 
strong interaction within the nucleus, a formidable job that was bypassed 
by Dover {\it et al.} \cite{DGR83} and in the present work using 
a well-constrained phenomenological $\bar n$-nuclear potential. 

Another source of uncertainty involves {\it two-nucleon} processes which 
inside the nucleus might compete with the leading one-nucleon process 
considered here. In their 1985 paper, making contact with beta-decay and 
EMC calculations, Dover {\it et al.} \cite{DGR83} estimated these additional 
modes of neutron disappearance to be about $15\%-30\%$, and largely incoherent 
with the one-nucleon mode. This provides a {\it systematical} uncertainty 
which may be used to {\it increase} the stated lower bound on 
$\tau_{n\bar n}$.

\section*{Acknowledgments} 

One of us (A.G.) thanks Yuri Kamyshkov for his support and hospitality 
during the International Workshop on B-L Violation, Berkeley, September 
2007, where a related preliminary presentation was made. We thank John 
Millener for useful correspondence on translationally invariant nuclear 
wavefunctions. This research was partially supported by the Israel Science 
Foundation grant 757/05.

\section*{Appendix: asymptotic $\bar n$ radial wavefunctions} 

Here we record special, regular at $r \rightarrow \infty$ solutions of the 
$\bar n$ inhomogeneous radial Eq.~(\ref{eq:nbarfinal}) in terms of similar 
solutions of the $n$ homogeneous radial Eq.~(\ref{eq:nfinal}). At $r>R$, 
where the $\bar n$ and $n$ nuclear potentials are negligible, these equations 
are written in standard form as 
\begin{equation} 
-v_{\nu{\ell}j}^{''}(\rho)+\frac{{\ell}({\ell}+1)}{\rho^2}v_{\nu{\ell}j}(\rho)
+v_{\nu{\ell}j}(\rho)+\frac{1}{B_{\nu{\ell}j}^{(n)}}u_{\nu{\ell}j}(\rho)=0~, 
\label{eq:nbarasympt} 
\end{equation} 
\begin{equation} 
-u_{\nu{\ell}j}^{''}(\rho)+\frac{{\ell}({\ell}+1)}{\rho^2}u_{\nu{\ell}j}(\rho)
+u_{\nu{\ell}j}(\rho)=0~, 
\label{eq:nasympt} 
\end{equation} 
for $\bar n$ and $n$, respectively, where $\rho=\kappa_{\nu{\ell}j}r$ is 
dimensionless. The $n$ bound-state, regular at $\rho \rightarrow \infty$ 
solutions are given by 
\begin{equation} 
u_{\nu{\ell}j}(\rho) = A_{\nu{\ell}j}{\cal P}_{\ell}(\rho)\exp(-\rho) = 
A_{\nu{\ell}j} (-1)^{\ell}~\rho^{{\ell}+1} 
(\frac{1}{\rho}\frac{\rm d}{\rm d \rho})^{\ell}~\frac{\exp(-\rho)}{\rho}~,  
\label{eq:nsolasympt} 
\end{equation} 
where $A_{\nu{\ell}j}$ are normalization constants ensuring that 
asymptotically ${\cal P}_{\ell}(\rho\rightarrow \infty) \to 1$.  
${\cal P}_{\ell}(\rho)$ are polynomials in $1/\rho$, related to the outgoing 
spherical Hankel functions \cite{ASt70}, satisfying the differential equation 
\begin{equation} 
{\cal P}_{\ell}^{''}(\rho)-2{\cal P}_{\ell}^{'}(\rho)-\frac{{\ell}({\ell}+1)}
{\rho^2}{\cal P}_{\ell}(\rho)=0~. 
\label{eq:Pdiff} 
\end{equation} 
The lowest-order ${\cal P}_{\ell}$s relevant to the present work are 
\begin{equation} 
{\cal P}_0(\rho)=1~,~~~{\cal P}_1(\rho)=(1+\frac{1}{\rho})~,~~~
{\cal P}_2(\rho)=(1+\frac{3}{\rho}+\frac{3}{\rho^2})~,~~~{\cal P}_3(\rho)=
(1+\frac{6}{\rho}+\frac{15}{\rho^2}+\frac{15}{\rho^3})~. 
\label{eq:Ps} 
\end{equation}  
A useful recursion relation satisfied by the ${\cal P}_{\ell}$s is 
\begin{equation} 
{\cal P}_{\ell}(\rho)=(1+\frac{\ell}{\rho}){\cal P}_{\ell-1}(\rho)-
{\cal P}_{\ell-1}^{'}(\rho)~~~~~~ ({\cal P}_{-1}(\rho) \equiv 1)~, 
\label{eq:Prec} 
\end{equation} 
easily derived from the explicit form of ${\cal P}_{\ell}$ given in 
Eq.~(\ref{eq:nsolasympt}). 

It can be shown that regular at $\rho \rightarrow \infty$ solutions of the 
$\bar n$ inhomogeneous radial Eq.~(\ref{eq:nbarasympt}) are given by 
\begin{equation} 
v_{\nu{\ell}j}(\rho) = -\frac{A_{\nu{\ell}j}}{2B_{\nu{\ell}j}^{(n)}}~\rho~
{\cal Q}_{\ell}(\rho)\exp(-\rho) 
~,  
\label{eq:Qs} 
\end{equation} 
where ${\cal Q}_{\ell}(\rho)={\cal P}_{\ell-1}(\rho)$. For a proof, 
one forms the inhomogeneous second-order differential equation satisfied 
by the ${\cal Q}_{\ell}$s, making use of Eqs.~(\ref{eq:Pdiff}) and 
(\ref{eq:Prec}). The latter equation allows us then to construct the 
${\cal Q}_{\ell}$s recursively: 
\begin{equation} 
{\cal Q}_{\ell+1}(\rho)=(1+\frac{\ell}{\rho}){\cal Q}_{\ell}(\rho)-
{\cal Q}_{\ell}^{'}(\rho)~~~~~~({\cal Q}_0(\rho)=1)~. 
\label{eq:Qrec} 
\end{equation}

\end{document}